\documentclass[prd,aps,twocolumn,showpacs,preprintnumbers,amsmath,amssymb,nofootinbib]{revtex4}
\usepackage[dvips]{graphicx} 
\usepackage{amsmath} 
\usepackage{amssymb} 
\usepackage{graphicx}
\usepackage{dcolumn}
\usepackage{bm}
\def\be{\begin{equation}} 
\def\ee{\end{equation}} 
\def\bea{\begin{eqnarray}} 
\def\eea{\end{eqnarray}} 
 
\begin{document} 
 
 
\date{\today} 
 
\title{Canny Algorithm, Cosmic Strings and the Cosmic Microwave Background} 
 
\author{Rebecca J. Danos and Robert H. Brandenberger 
\email[email: ]{rjdanos,,rhb@hep.physics.mcgill.ca}}
 
\affiliation{Department of Physics, McGill University, 
Montr\'eal, QC, H3A 2T8, Canada}

\pacs{98.80.Cq} 
 
\begin{abstract} 

We describe a new code to search for signatures of cosmic strings in cosmic microwave
anisotropy maps. The code implements the Canny Algorithm, an edge detection algorithm 
designed to search for the lines of large gradients in maps. Such a gradient signature
which is coherent in position space is produced by cosmic 
strings via the Kaiser-Stebbins effect. 
We test the power of our new code to set limits on the tension of the cosmic strings by
analyzing simulated data with and without cosmic strings. We compare maps with a
pure Gaussian scale-invariant power spectrum with maps which have a contribution
of a distribution of cosmic strings obeying a scaling solution. The maps have angular
scale and angular resolution comparable to what current and future ground-based
small-scale cosmic microwave anisotropy experiments will achieve. We present
tests of the codes, indicate the limits on the string tension which could be set with
the current code, and describe various ways to refine the analysis.  
Our results indicate that when applied to the data of ongoing
cosmic microwave experiments such as the South Pole Telescope
project, the sensitivity of our method to the presence of cosmic
strings will be more than an order of magnitude better than
the limits from existing analyses.

\end{abstract} 
 
\maketitle

\newcommand{\eq}[2]{\begin{equation}\label{#1}{#2}\end{equation}} 
 
\section{Introduction} 

There has been a recent renaissance of interest in cosmic strings 
(see e.g \cite{recentCS} ) as a mechanism contributing to the 
spectrum of primordial cosmological perturbations and the corresponding
anisotropies in cosmic microwave anisotropy (CMB) maps. In part
this renewed interest is sparked by the realization that many supergravity
models of inflation also lead to the production of strings at the end of the
period of inflation \cite{Rachel}. In addition, many models of inflation
in the context of string theory predict the formation of a network of strings
at the end of the inflationary phase \cite{CS-BI}. The
conditions under which these cosmic
superstrings \cite{Witten} are stable have been explored in \cite{Pol1}.
Cosmic superstrings may also arise in other approaches to superstring
cosmology such as the Ekpyrotic scenario \cite{Ekp} or string gas
cosmology \cite{BV,NBV}.

If matter is described by a particle physics model which admits stable strings
forming after inflation, then by causality it is inevitable that a network of such 
strings will form during the cosmological phase transition which corresponds 
to the symmetry breaking which is responsible for the existence of the strings
\cite{Kibble}. It is inevitable that the network of strings contains ``infinite"
strings (strings crossing the entire volume). For applications in cosmology
one divides the strings into ``loops" (loops of cosmic string with a curvature
radius smaller than the Hubble radius $H^{-1}(t)$, where $H(t)$ is the
cosmological expansion rate) and ``long strings" (infinite strings and
loops with curvature radius larger than the Hubble radius). The 
causality argument \cite{Kibble} (see also 
\cite{VS,HK,RHBrev} for some standard reviews on cosmic strings and
structure formation, and \cite{original} for the original
references) implies, in fact, that at all times after the phase transition
a network of long strings with a correlation length (mean curvature radius) 
comparable or smaller than
the Hubble radius will survive. It can be argued 
\cite{Vilrev,VS,HK,RHBrev} 
that a network of (non-superconducting) strings will approach a ``scaling solution"
for which at all late times $t$ the correlation length of the network of
long strings is a fixed fraction of the Hubble radius 
\footnote{It is likely that the distribution of cosmic string loops also approaches
a scaling solution; see e.g \cite{AT,BB,AS} for early simulations of
cosmic string networks supporting the conclusion that the distribution
of loops also scales, \cite{Hind} for an opposing view based on
field theory simulations and \cite{Van,Mart,Ring,Shell,Pol2} for more recent work
supporting scaling for string loops.}.

Cosmic strings \cite{NO} 
are described by their mass per unit length $\mu$ which is
usually quoted in terms of the dimensionless number $G \mu$, $G$
being Newton's gravitational constant. A straight string has
a tension which is equal in magnitude to $\mu$. Hence, the
effective action which describes the motion of a straight string
is the Nambu-Goto action, the same action which describes
fundamental strings. Since cosmic strings carry
energy, they will lead to density fluctuations and
cosmic microwave background (CMB) anisotropies. 

The network
of cosmic strings will generate a scale-invariant spectrum of
cosmological perturbations \cite{CSstructure}. More relevant to the
current paper, cosmic strings generate a very specific signature in
cosmic microwave background anisotropy maps, namely line
discontinuities \cite{KS}. These line discontinuities arise
since space perpendicular to a cosmic string is a cone with
deficit angle given by \cite{deficit}
\be 
\alpha \, = \, 8 \pi G \mu \, .
\ee
Since the motion of cosmic strings is relativistic,  photons passing on 
different sides of the string moving with a velocity $v$ perpendicular
to the plane spanned by the string direction and the line of sight
between the observer and the string will be seen by the observer
with a relative Doppler shift
\be \label{KSsig}
{{\delta T} \over T} \, = \, 8 \pi \gamma(v) v G \mu \, ,
\ee
where $\gamma(v)$ is the relativistic gamma factor (see Figure 1). 
Looking in direction of the string, we will see a line in the sky across 
which the CMB temperature jumps by the above amount. We will
denote this effect, the Kaiser-Stebbins effect, as the KS effect in the rest of the paper. 

\begin{figure}
\includegraphics[height=6cm]{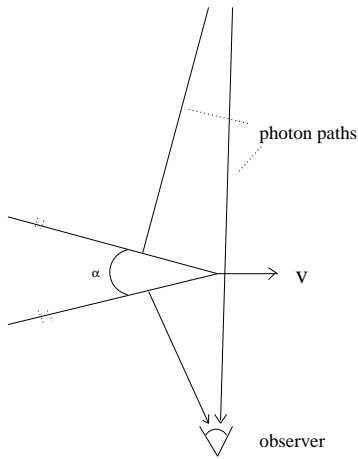}
\caption{Geometry of the Kaiser-Stebbins effect: Photons passing
on the two sides of the moving cosmic string obtain a relative Doppler
shift for the observer who is at rest.} \label{fig:1}
\end{figure}

The conical deformation of space induced by a ``long" cosmic string
has a length of the order the Hubble radius, i.e. of the order of $t$ in
direction of the string. Since cosmic strings are formed in a phase
transition and the effects of the transition travel with the speed of
light, the depth of the region affected by the string (in direction perpendicular
to the string) is $t$, as shown in \cite{Joao}. Hence, each string 
between $t_{rec}$ and $t_0$ whose
``zone of influence" \footnote{By which we mean the region of space which
is deformed due to the presence of the string.} is intersected by the past
light cone of the observer will induce a line discontinuity in the CMB
temperature map with length $g t$, where $g$ is a random number in the
interval $0 < g < 1$ which takes into account the random angle of the 
velocity vector of the string relative to the plane determined by the string
direction and the observer's line of sight to the string. The anisotropy
pattern induced by a single string segment contains several edges.  There is
the central line discontinuity (\ref{KSsig}) 
coming from photons which pass on
different sides of the string.  Since the deficit angle in Figure \ref{fig:1}
has finite depth (in direction $-v$), as explained in the previous paragraph
there will be two edges (again sharp because the deficit angle sharply
decreases to zero \cite{Joao}) with half the value of $\delta T$.  Finally, our
modelling of the string network in terms of straight segments introduces the
sharp ``boundary'' edges perpendicular to the central edge (see also
Figure \ref{teststringmap}).

To detect the line discontinuities in CMB maps produced by
cosmic strings, it is important to have small angular 
resolution. Strings present in the universe between the time of
last scattering and the present time contribute to the signal.
However, according to the cosmic string scaling solution, the most
numerous strings are those present at or shortly after last
scattering. The Hubble radius at that time subtends an angle
of about $1.8^{o}$. If the angular resolution is not much smaller
than this angle, then the anisotropies produced by these strings will
not be distinguishable from anisotropies produced by Gaussian noise
with the corresponding coherence length. On the other hand, full
sky coverage is not essential. Thus, ground-based small angular resolution
surveys such as ACT \cite{ACT} or the South Pole Telescope \cite{SPT}
both of which have angular resolution of about $1^{'}$ are ideal
to search for strings. The Planck satellite experiment \cite{Planck} with an angular
resolution of about $5^{'}$ will also yield a good data set to use,
in particular since the systematic errors in the data will likely
be smaller.

The KS effect is a part of the ``Integrated Sachs-Wolfe" \cite{SW}
contribution to CMB anisotropies. The primordial cosmological fluctuations 
produced by strings also contribute to the regular Sachs-Wolfe
effect. In contrast to cosmological fluctuations produced in inflationary
cosmology, those produced by cosmic strings are ``incoherent" and
``active" as opposed to ``coherent" and ``passive" \cite{incoherent}. 
The string network is continuously seeding growing curvature fluctuations 
on super-Hubble scales, and therefore the fluctuations enter the Hubble 
radius not as standing waves. As a consequence, the angular power spectrum of
CMB anisotropies does not have \cite{Periv,Albrecht,Turok}
the acoustic ringing associated with
coherent passive fluctuations \cite{SZ}. Since acoustic ringing
has been observed with recent high precision CMB measurements
\cite{Boomerang,WMAP}, it is now clear that cosmic strings cannot
be the main source of cosmological fluctuations. Their contribution is
bounded from the accurate measurements of the angular power
spectrum of the CMB in the region of the first acoustic peak to be
less than $10\%$ \cite{CMBlimits} which corresponds to a value
of $G \mu$ of about $3 \times 10^{-7}$. In the literature, one finds
slightly stronger constraints which come from pulsar timing data
\cite{Pulsarlimit,timinglimits}. However, pulsar bounds make use of estimates
of the spectrum of gravitational radiation from cosmic string loops.
Since there is still a lot of uncertainty about the distribution of cosmic
string loops, and since the amount of gravitational radiation from
a fixed loop is also rather uncertain, bounds on $G \mu$ coming
from millisecond pulsar timing are not very robust. Much more robust
signatures come from the long and straight strings,
signatures which we are discussing in this paper. 

Early work to identify the KS signal of cosmic strings in CMB anisotropy
maps was presented in \cite{Moessner} which concluded that the
angular resolution of WMAP would not be small enough to resolve
the KS signature. After the release of the WMAP data \cite{WMAP},
there were two sets of analyses introducing new algorithms to
look specifically for the KS signature. Lo and Wright \cite{Lo}
applied a matched filtering method, whereas Jeong and Smoot
\cite{Smoot} introduced new statistics such as one measuring the
connectedness of neighboring temperature steps or another
one proposing a decomposition of the temperature map into
constant, Gaussian and straight string step components. Both
groups applied their statistics to the WMAP data. From the
null results of the searches, a direct upper bound on the string
tension of $G \mu < 10^{-6}$ could be set, a bound not competitive
with the existing bounds from the matching of the angular power
spectrum.

In \cite{ABB}, it was proposed to make use of the Canny
algorithm \cite{Canny} to search for the KS effect, and the
preliminary analysis showed that the statistic offers the
promise to improve the limit on $G \mu$ from direct
searches for the KS signal by a large factor. The Canny
algorithm is an edge detection algorithm which was
previously used in image recognition work and metallurgy.
It looks for lines in a map across which there is a large
gradient. Thus, the algorithm appears to be well suited to
detect the KS signature. In this paper we present a new
and improved implementation of the Canny algorithm and
apply it to test data. In agreement with \cite{ABB} (and
with the followup paper \cite{Stewart}), we
find the new analysis based on the Canny algorithm is
able to find or rule out strings with a tension greater
than an upper bound which is more than an order of
magnitude smaller than previous bounds derived from
looking for the KS signature of strings.
 We emphasize
that the code on which this paper is based was
developed completely independently from that used
in \cite{ABB} and \cite{Stewart}. It is different in
structure and in fact is based on a different programming
language. The fact that the results reported here
agree with those in \cite{Stewart} presents a very
important check on the methods.

The Canny algorithm works in position space. Starting
from an anisotropy map, it first produces an ``edge map'',
the edges corresponding to lines in the sky 
perpendicular to which the
gradient of the anisotropy map is sufficiently large
\footnote{As detailed in Section 3, a new aspect of the
present code is to search for gradients in a range tuned
to the expected KS signal.}. The edge detection algorithm
must be able to take into account the fact that
Gaussian noise superimposed on top of the cosmic string
signal will produce large variations in the magnitude
of the gradient along a cosmic string edge. Given the
edge map, a second algorithm counts the number of edges
of fixed length and produces a histogram of edge lengths.
Both the edge detection and the edge counting algorithms
contain various parameters and thresholds which can
be set by the user and which can be tuned to give
the algorithm maximal discriminatory power, and are
improved over the original code presented in \cite{ABB}. 
The values of the thresholds and parameters will depend on 
the specific data. The final step of the code is a
statistical comparison between the histograms produced
in the previous step for data with and without
cosmic strings.

The outline of this paper is as follows: In the following
section we discuss the construction of the test data maps -
both the pure Gaussian maps and the maps containing 
cosmic strings. In Section 3 we present the new implementation
of the Canny algorithm. We first describe the algorithm which takes the
CMB temperature map and converts it into an edge map. Then, we
turn to the separate algorithm which is used to produce a histogram of edge
lengths. Section 4 presents the results from the application of
the Canny algorithm to the test maps. We conclude with a summary 
of the method and results, focusing on refinements of the code
which can be made to improve the code's discriminatory power,
and give a preview of future applications.

\section{Simulations}
\label{simuls}

In this section we describe the codes which were used to create
the simulated temperature maps. The maps have two components, firstly
a Gaussian map with an angular power spectrum corresponding to
the inflationary ``concordance model",  the second a map of anisotropies
produced by long cosmic strings according to the KS effect.
These simulation routines have been created from scratch, without
using any input from the existing code \cite{ABB}. The simulation
routines are written in Interactive Data Language (IDL).
Both for the inflationary fluctuations and the cosmic string maps, the
theory predicts ensemble averages from which particular classical
realizations are drawn. To obtain firm predictions, a large number of
simulations must therefore be run.

Since we have in mind applying our code to surveys with
small angular resolution but only partial sky coverage, we
work in the ``flat sky'' approximation \cite{White} in which
a segment of the sky is approximated by a rectangle. This
approximation is made because of computational ease, since
the basis solutions of the wave equation in flat space, the
Fourier modes, are much easier to work with than the basis
functions on a sphere, the spherical harmonics.

\subsection{Gaussian Map}

First we create a square grid temperature
map of the Gaussian inflationary perturbations. The grid size is
set by the angular resolution and by the angular size of the survey
which we want to simulate (the number of grid points along
an axis is $N_{max} = L / R$, where $L$ is the extent of the survey
along the axis being considered, and $R$ is the angular resolution). 
The map is constructed based on the angular power spectrum of
CMB fluctuations computed using one of the standard codes
used in the literature (see below).

A differential temperature value, $\Delta T/T$, (corresponding to the 
temperature with the monopole subtracted) is assigned to each grid 
point $(n, m)$ ($n$ and $m$ are integers) of the map, where 
each grid point represents a position in the window of the observed 
sky.  If $L$ is the length of a side of the window of observation in degrees
in the first coordinate direction,
then the angular distance of a grid point $(n, m)$ from the edge is
$x = nL/N_{max}$. The two dimensional vector is designated by
$\bm{x}$.

The code is designed to construct a square grid. If rectangular areas are 
needed, we can construct a larger square temperature map and cut out the
appropriate rectangle from this larger grid.  
The applicable scales are less than $60^{\circ}$ indicating the applicability of 
the ``flat sky'' approximation. 

In general, the temperature anisotropy of the CMB is expressed in terms of
spherical harmonics $Y_{lm}(\theta, \phi)$, $\theta$ and $\phi$ being latitude
and longitude, respectively:
\be
\frac{\Delta T}{T} (\theta, \phi) \,  = 
\, \sum_{l,m} a_{lm}Y_{lm} (\theta, \phi) \,
\ee
where $a_{lm}$ are the coefficients of the expansion.
However, since the observed window will be less than 60 degrees (we will
take it to be approximately 10 degrees), plane waves can be substituted for
$Y_{lm}$ per the flat sky approximation:
\be
\frac{\Delta T}{T}(\bm{x}) \, = \, \sum_k T(\bm{k})e^{i\bm{k}\cdot\bm{x}}.
\ee
(N.B. In the following equations $T(\bm{x})=\Delta T(\bm{x})/T$ and
$T(\bm{k})=\Delta T(\bm{k})/T$)
Comparing these equations, it is obvious that the coefficients $T(\bm{k})$
correspond to $a_{lm}$ so that:
\be \label{anglerel}
\langle T(\bm{k})^2\rangle \, = \, \langle a_{lm}^2 \rangle \, 
\equiv \, C_{l(k)},
\ee
in which the angular brackets stand for ensemble averaging, and the
$C_{l}$ give the angular power spectrum of the CMB. According
to the ergodic hypothesis, the ensemble average is equivalent to the
spatial average. Thus,
\be
\langle a_{lm}^2\rangle \, = \, \frac{1}{2l +1}\sum_{m=-l}^l a_{lm}^2,
\ee
which, as indicated in (\ref{anglerel}) is the square of the width of a Gaussian 
distribution, namely $T(\bm{k})$.    
The probability of a given $T(\bm{k})$ follows a Gaussian
distribution just as the $a_{lm}$  follow a Gaussian distribution.

The temperature map in Fourier space, $T(\bm{k})$, is arranged into
a grid in which each grid point is assigned physical angular
coordinate values $k_{x}$ and
$k_{y}$, which are related to the numerical $k_{xn}$ and $k_{yn}$ values
(each ranging between 0 and $N_{max}-1$) by
\bea
k_{x} \, &=& \, \frac{2\pi}{L}(k_{xn}-k_{max})\\
k_{y} \, &=& \, \frac{2\pi}{L}(k_{yn}-k_{max}) \nonumber
\eea
in which the $k_x$ and $k_y$ values run from $-k_{max}$ to $k_{max}$. The
value of $k_{max}$ corresponds to the angular resolution of the
data set. Note
that this setup requires us to have an odd number of pixels in our
Gaussian map.  If an even number of pixels is required, the Gaussian map
can be stripped of a row and column once it is constructed.
  
Using
\be
\lambda \, = \, \frac{360^\circ}{l} \, = \, \frac{2\pi}{k},
\ee
in which the wavelength, $\lambda$ is measured in degrees and the wavenumber
$k$ is measured in inverse degrees,
we compute the dependence of the degree of the spherical harmonic 
(mode number) $l(k)$ on the magnitude $k$ of $\bm{k}$.  
Once this is done, we can look up the
corresponding $C_l$ for each $\bm{k}$ from the output of a
standard CMB simulation code (see below).  
Since $l(k)$ is not necessarily an integer, for each value of $k$, 
the program then determines the value $C_l(k)$ by linear
interpolation between the $C_l$ values of the two closest values
of $l$  corresponding to each $l(k)$.

Given the $C_l$ values determined as described above, the 
Fourier space temperature map $T(k)$ can be determined by
\be
T(k_x,k_y) \, = \, \sqrt{C_{l(k_x,k_y)}/2}(g_1(k_x,k_y)+ig_2(k_x,k_y))
\ee
where $g_1(k_x,k_y)$ and $g_2(k_x,k_y)$ are 
randomly generated numbers obtained from a Gaussian distribution
with variance one and mean of zero.

Since the position space temperature map $T(\bm{x})$ must be real, 
then $T(\bm{k})$ must satisfy
$T(\bm{k})=T^*(-\bm{k})$.  

\begin{figure}
\includegraphics[scale=0.8]{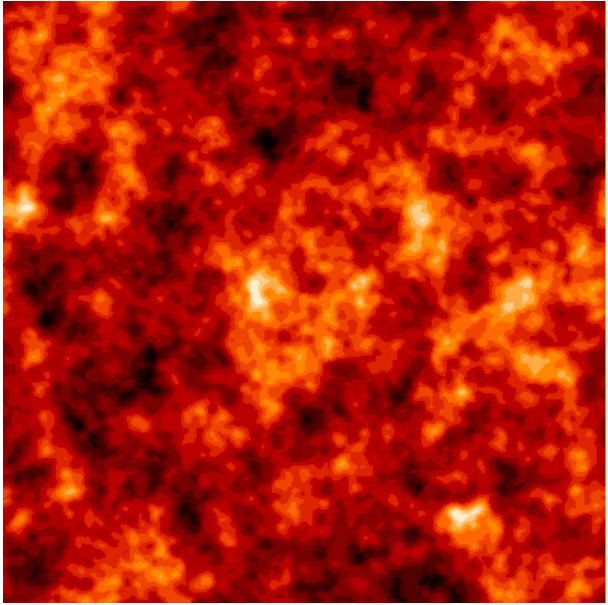}
\caption{Figure of the gaussian temperature map with a window of $10^{\circ}$
and a resolution of 1.5 arcminutes.}
\label{fig:testgaussianmap}
\end{figure}

In summary, we derive $T(\bm{k})$ from the values of $C_l$ determined as
described above, and $T(-\bm{k})$ making use of the Hermitian property.
Finally, to compute $T(\bm{x})$, we perform a Fast Fourier Transform on 
$T(\bm{k})$.
Figure (\ref{fig:testgaussianmap}) shows the output of a temperature map
with pure Gaussian noise.

We use the Code for Anisotropies in the Microwave Background (CAMB)
\cite{CAMB} to
generate the CMB temperature map generated by inflationary perturbations.
CAMB, like the code on which it is based, CMBFAST \cite{CMBFAST}, 
employs the
line of sight integration prescription to compute CMB anisotropies.  The 
advantage of using CAMB is that it computes $C_l$'s for higher $l$'s.  Whereas
CMBFAST can at best compute the power spectrum to $l=3000$, corresponding to
a resolution of 10 arcminutes, we use a power spectrum generated by CAMB up
to $l=22000$, corresponding to a resolution of 1.39 arcminutes.  Simulations
with higher resolution will enable us to place lower bounds on the
string tension of cosmic strings.  Additionally, high resolutions are 
necessary to apply our algorithm in the future high resolution CMB
experiments such as ACBAR \cite{ACBAR} and the South Pole Telescope \cite{SPT}.

The parameters we use for our simulations are standard $\Lambda$CDM 
concordance parameters based on fitting the model to
a multi-experiment data set  CMBall \cite{arXiv:0801.1491}, namely
$\Omega_bh^2 = 0.0227$, $\Omega_ch^2 = 0.112$, $\Omega_m=0.26$, 
$H_0 = 71.8$, $n_s = 0.965$, $w=-1$ (the equation of state for dark energy), 
$\Omega_{\nu}h^2=0$, $T_{CMB}=2.726$ (the temperature of the CMB), 
$Y_{He}=0.24$, and reionization optical depth, $\tau_{reion}= 0.093$.

\subsection{Temperature Map from pure Cosmic Strings}
 
Next we generate a temperature map
of superimposed straight line segments representing the
effects of long cosmic strings obeying  the scaling solution.  
As in the Gaussian 
simulation, each grid point is assigned a temperature 
fluctuation $\Delta T/T$.

For simplicity we do not include the effects of string loops or
of the small scale structure on the strings. According to the current status of
cosmic string simulations, the effects of long strings dominate the 
cosmic string CMB sky. Small-scale structure on the long strings
would maintain the KS effect (albeit with slightly reduced amplitude).
Thus, in terms of searching for the KS signature of strings, the simplifications
we use can be justified as a first step. Going beyond this approximation
would require much more involved simulations (like those of \cite{Fraisse})
with many more parameters. 
 \footnote{Although in
principle cosmic string models are determined by a single parameter,
namely $\mu$, the cosmic string evolution cannot be
followed either analytically or numerically without making additional
assumptions, in particular on how cosmic strings interact. The results
obtained in different numerical analyses concerning the string loop
distribution vary by orders of magnitude (see e.g. the different results
obtained in \cite{AT,BB,AS}), whereas the results for the
distribution of long strings are comparable. Thus, whereas limits
based on long strings are robust, those which include effects of
string loops are not.}
In future work we plan to apply our
Canny algorithm code to temperature maps obtained from more
sophisticated simulations
that include more detailed structure of cosmic strings.

In contrast to the Gaussian string map which is set up in Fourier space
and transformed to position space via a Fourier transform, the
string maps are set up directly in position space. As explained in the
Introduction, each long string whose region of influence intersects
the past light cone of the window in the sky we are considering will
contribute a KS signal.
We assume that the area in the sky affected
by the temperature discontinuity of a single straight string 
segment per the KS 
effect is a rectangular box on each side of the string. The depth of
the box (length perpendicular to the string) is
equal to the Hubble length \cite{Joao}. The size of the box
in direction of the string is taken to be the product of the Hubble length
and a length coefficient $\gamma$. The Hubble  length corresponds to the
time at which the past light cone intersects the region of influence
of the string. We apply the KS signature by adding a positive
temperature fluctuation to grid points in the box on one side of the string,
and by adding the corresponding negative temperature  
fluctuation to the grid points in the box on the other side of the string.  

We are using a toy model for the cosmic string distribution first introduced
in \cite{Leandros} and also used by Pogosian et al. in \cite{CMBlimits}. 
It is based on a simple
representation of the distribution of long strings as described by the
scaling solution. According to this solution, the distribution of long strings is
at all times like a random walk with a step length which is of the order of
the Hubble radius at that time. We will denote this length by
$\gamma H^{-1}$, where $\gamma$ is a constant factor of order
one. Physically, this length represents the mean curvature radius of
the string network. In the toy model, this distribution is approximated
by a collection of straight string segments of length $\gamma H^{-1}$.

One can argue that the curvature radius of the network of long strings
must be of the order of the Hubble radius.
If the curvature radius is larger than the Hubble radius, the strings will 
sit there as the universe expands until the Hubble radius catches up.  If the
curvature radius is less than the Hubble radius, then the strings will
oscillate wildly, intersect and split off string loops (which then decay
by gravitational radiation) until the curvature radius catches up to the
Hubble radius.  So the 
curvature radius will always be of order of the Hubble radius.  The key
ingredients in this argument are, firstly, that the effective action for a cosmic
string is the Nambu-Goto action and hence the string dynamics is
relativistic, and secondly that when strings cross they will intercommute, i.e.
exchange ends, thus allowing the production of string loops.  

Since the strings are relativistic, the distribution of strings will have changed
after one Hubble time. Snapshots of string distributions at time steps
separated by Hubble time intervals will, when rescaled to the Hubble
radius, look like independent realizations of a stochastic process.
Hence, in our toy model of the distribution of long strings, we take the
string segments to be independently distributed over time intervals
greater than a Hubble time.

The input parameters for the string map generating function 
are the number of pixels on each side
of the window, the number of degrees the window subtends, the number of
strings per Hubble volume per the scaling solution, the string tension,
$G \mu$, and the length coefficient $\gamma$.

The procedure to simulate cosmic strings is to follow the past light cone
in time intervals $t_{n-1} \leq t < t_n$, where $t_{n+1} = \alpha_1 t_n$,
from the time of last scattering to the present time. 
If $\alpha_1=t_{n+1}/t_n=e$  then there are approximately 15 time intervals.

Next we project all strings present in one Hubble time interval 
to a fixed time hypersurface at the center of the time interval. This is done
for all time intervals. On the microwave sky, the Hubble length
in degrees corresponding to the comoving distance of the Hubble radius
in space at the n'th time interval  can be shown to be 
(making use of the equation of state and of the Friedmann equation)
\be
d_c(t_{n+1}) \, = \, d_c(t_{n})\alpha_1^{1/3} \, ,
\ee
in which $d_c(t)$ is in degrees.  This recursion relation starts with 
$d_c({t_0}) = 1.8$.

We need to find all of the strings which for a fixed time interval influence
the CMB temperature map of the observed window.
Instead of simulating only strings in the observed window, we must
consider all strings in an extended window. This
protects us from missing strings that start outside of the observed window
but extend into it.  At each time interval, the extended window
is a square with each side subtending the degrees of the observed window added
to twice the Hubble distance for that time step.  The observed window is the
central section with one Hubble length on either side.

Next we compute the number of Hubble volumes the extended window
subtends at each time step, and the number of strings in each
extended window.  The number of Hubble volumes is obtained
as the square of the number of 
degrees subtended by each extended window divided by
the degrees subtended by each
Hubble distance.  The number of strings expected in the 
extended window for each time step
is the rounded product of the strings per Hubble volume (determined as an input
parameter based on the scaling solution) and the fraction/number of Hubble 
volumes a window subtends for each of the 15 time intervals.

Next we loop through each time step and each expected string in the extended
window to determine where the string should be placed in the extended window
and to place it by constructing a temperature differential.  This is completed
in a separate function in the program.

To simulate the strings, the program randomly places a line segment with
a length given by $\gamma \cos\alpha H^{-1}$, where the Hubble length 
$H^{-1}$ is in radians for the given time step. The angle $\alpha$ runs from
0 to $\pi/2$.  Multiplying by $\cos\alpha$ accounts for a projection from
three to two dimensions.  Although technically the length of the string should
be multiplied by a coefficient given by projecting a geodesic on a sphere
onto a plane, since we are dealing with small window sizes and the flat sky
approximation is appropriate, this complication is unnecessary.  The location 
of the beginning of the string and its direction are determined randomly.
The location of the beginning of the string is
given by taking the product of a random number 
between 0 and 1 for both directions on the grid and the length of the
extended window in radians.  The direction of the placement of the string, i.e.
the angle $\theta$ of the string, is chosen by a randomly generated number 
between $-\pi/2$ and $\pi/2$.  This is the direction of the string as given
by the angle from the x-axis.  Finally,
a binary flag is used to determine which side of the string will be a positive
temperature fluctuation and which side will be a negative temperature
fluctuation.  The value of the flag is randomly determined.

Next the code computes (from the location of the start of the string and the
direction of the string) the end point of the string and each corner of the
box around each side of the string and each slope for each boundary of the
box. 

The routine loops over each pixel in the observed window, starting in the lower
left hand corner, to determine if the pixel lies within the box affected
by the cosmic string.  If it does, the temperature is changed by $\delta T$.
The sign of $\delta T$ is determined by the temperature flag delineated above.
The temperature fluctuation is given by
\be
\frac{\delta T}{T} \, =  \, \tilde{v}r4\pi G\mu \, 
\ee
in which $T=2.726$ is the background CMB temperature. In the above,
$\tilde{v}$ represents the root mean square (over all strings)
value of $v\gamma(v)$, where $v$ is the transverse velocity of the string
and its relativistic $\gamma$ factor is $\gamma(v)$. Also, 
$r$ a random number between 0 and 1 which adjusts the velocity to take into 
consideration the different velocities that the string might have as well
as the projection of the velocity of the string onto the plane perpendicular
to the line of sight. Based on recent cosmic string evolution simulations
we use the value $\tilde{v} = 0.15$.

\begin{figure}
\includegraphics[scale=0.8]{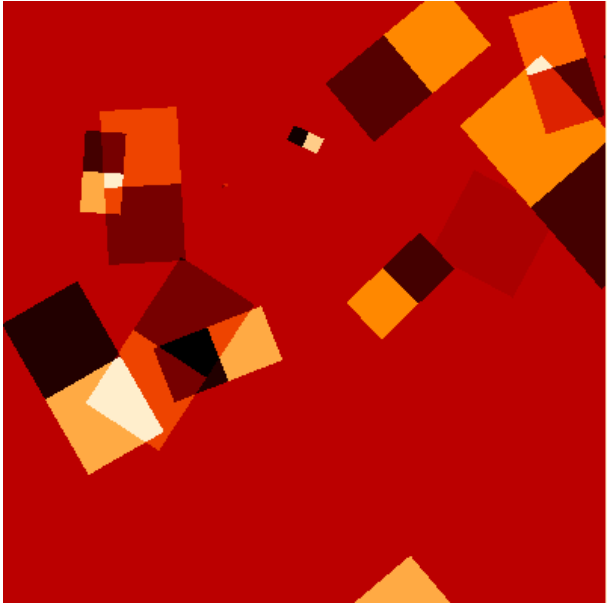}
\caption{A cosmic string simulation for a few strings.  The light areas 
represent where the temperature fluctuation is positive and the dark
areas represent where the temperature fluctuation is negative.}
\label{teststringmap}
\end{figure}

\begin{figure}
\includegraphics[scale=0.8]{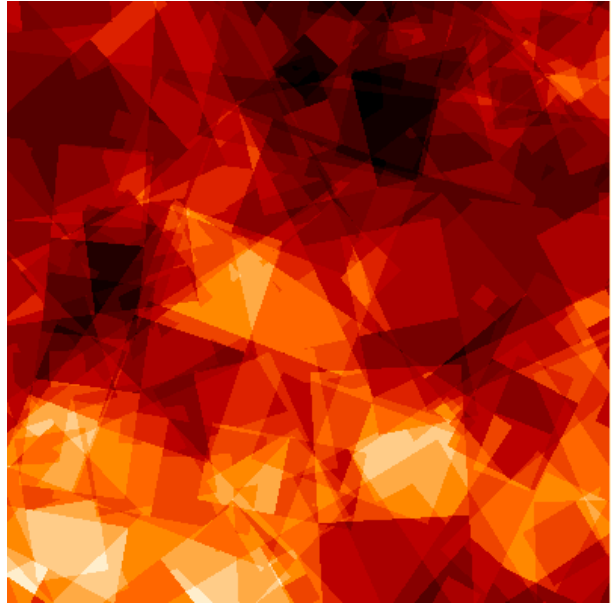}
\caption{The CMB temperature anisotropy map produced by a
scaling cosmic string simulation with $N = 1$. The discontinuity
lines in the maps produced by the KS effect are clearly visible,
but there are a lot of overlap regions where a number of
strings affect the temperature at a fixed point.}
\label{stringmapN=1}
\end{figure}

\begin{figure}
\includegraphics[scale=0.8]{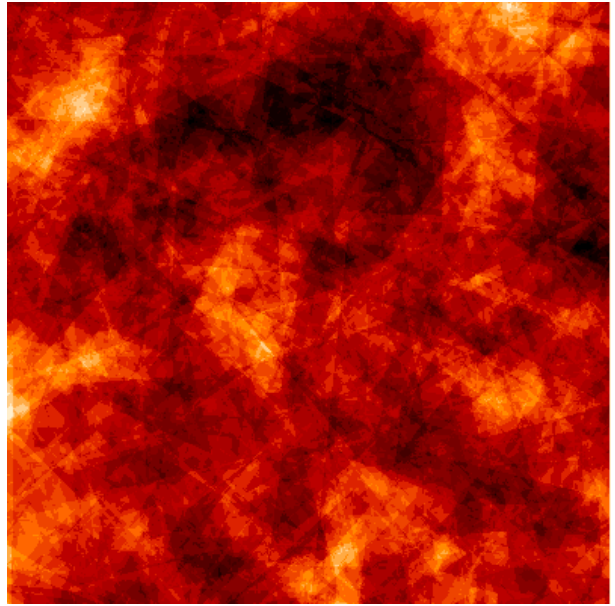}
\caption{The CMB temperature anisotropy map produced by a
scaling cosmic string simulation with $N = 10$. In this case
the effects of overlaps is much more pronounced.} 
\label{stringmapN=10}
\end{figure}

Figure \ref{teststringmap} shows the temperature map produced by
a simulation with a few test cosmic strings (not a scaling solution).
The temperature boxes produced by the individual strings are
clearly visible.
Figure \ref{stringmapN=1} demonstrates the corresponding results from
a full string simulation with $N=1$, i.e. one string
per Hubble volume.  The straight line temperature discontinuities
produced by individual strings are still visible. However, there are
a lot of overlap regions since a given photon will during its trajectory
pass close to several strings. The overlapping problem gets worse for
$N = 10$, as clearly visible in Figure \ref{stringmapN=10}.

\begin{figure}
\includegraphics[scale=0.5]{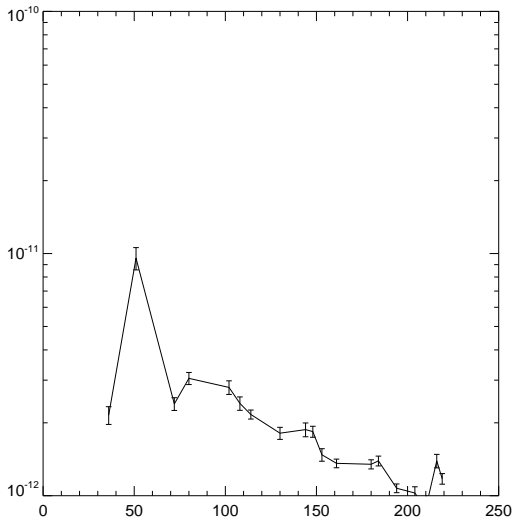}
\caption{The angular power spectrum of the CMB anisotropy
maps of pure cosmic string simulations with values $N = 10$
and $G \mu = 6 \times 10^{-8}$.
The horizontal axis is $l$, the vertical axis
is $l(l + 1) C_{l}$.} 
\label{Clspectrum}
\end{figure}

As a further test of the algorithm described in this subsection we
show the resulting angular power spectrum of temperature
anisotropies. The figure 
(Figure \ref{Clspectrum}) 
is for a pure cosmic string map
with $N = 10$ and $G \mu = 6 \times 10^{-8}$. The range of
$l$ values are limited from below by the angular size of
the simulation box, and from above by the angular resolution
we have chosen. The error bars are standard errors of the
mean based on $100$ runs. As theory predicts \cite{CSstructure,TTB},
the angular power spectrum is approximately scale-invariant
on large angular scales (the deviation from scale-invariance at
the lowest values of $l$ is presumably a boundary effect), 
and its amplitude is consistent with what is expected.

\subsection{Sum of Gaussian Map and Cosmic String Map}

Our goal is to test if the Canny algorithm is able to pick out the KS
signature from strings even if the strings are a subdominant component
to the CMB fluctuations. To test this, we need to produce temperature
maps which contain both cosmic strings for some value of $G \mu$
and a spectrum of Gaussian fluctuations like in the concordance
$\Lambda$CDM model, except with an amplitude of the Gaussian
noise which is reduced such that the total angular CMB power spectrum
remains consistent with either simulations or observations.

Here we describe how to find the coefficient, 
$a$, for each map
which will be used to add the pure string map to the pure Gaussian map
to get a Gaussian map with superimposed strings,
\be
T_{G+S}(\bm{k}) \, = \, aT_G(\bm{k}) + T_S(\bm{k}),
\ee
with $T_{G+S}$ as the temperature of the Gaussian map plus pure string map
and $T_G$ and $T_S$ as the temperature maps of the 
concordance Gaussian and string simulations respectively.  Alternatively,
we could use the temperature map of the best current data, the five         
year results of WMAP \cite{WMAPfive}, for $T_G$.  It is important
to use consistently either the temperature map from experimental data or
from theoretical simulations for both $T_G$ and $T_{G+S}$ since the $C_l$'s
for the two models are not identical. 

We want the model to give the best possible agreement with the data, either
experimental or simulated.  Hence,
the plan is to adjust the coefficient $a$ so that the combined
string and Gaussian map, $T_{G+S}(\bm{k})$, fits the data best.  This
means that we adjust the Gaussian map so that the corresponding 
$C_l$'s of  $T_{G+S}(\bm{k})$ give an
optimal fit to the data.  
Since the error bars on the observed angular power spectrum $C_l$
are smallest relative to the signal in the range
from $l_{min}=10$ to $l_{max}=220$ 
we shall fit the $C_l$'s within this range.  Note that this range includes 
most of the first Doppler peak region.  Now for values of $l$ comparable or
larger than that corresponding to the first Doppler peak, the contribution from
strings is dominated not by the Kaiser-Stebbins effect from strings between
$t_{rec}$ and $t_0$, but from string-induced fluctuations at last scattering
which are not included in our analysis.  
 Thus, there is an intrinsic inaccuracy in the determination of the
value of $a$. If we were to take $l_{max}$ to be the value where
the Kaiser-Stebbins effect ceases to be dominant (a value much
smaller than $220$), then we would obtain a smaller value for $a$
and hence a better discriminatory power of our algorithm. However,
this procedure would be worse than the one we have adopted,
since we would be working with a power spectrum which is a much
worse fit to the observations in the Doppler peak region than the
one we are using. Our choice of $l_{max}$ can thus to be
considered to be a conservative one.

The first step is to find the $C_l$ values from the pure string map.
To do this, the routine initially takes the inverse FFT of the temperature 
fluctuation $T(\bm{x})$ to compute $T(\bm{k})$.  Then, using $k = 2\pi l/360$, 
the routine computes the wave number magnitudes $k(l)$ for all
values of $l$ in the our range.  Then the value of $\delta k$   
corresponding to $\delta l=1/2$ is found.   

Next, the map summation routine finds the 
wavenumber magnitude $k$ for all values of $\bm{k}$ (whose
components range from $-k_{max}$ to $k_{max}$).  For each value of $l$
in the range between $l_{min}$ and $l_{max}$
the code finds the components 
$k_{x}$ and $k_{y}$ of all waves for which the magnitude of $k$
lies within $\delta k$ of $k(l)$. Knowing $k_{xn}$ and $k_{yn}$
we can find the corresponding $T(\bm{k})^2$ for each of the wavenumbers. 
Once all of the $T(\bm{k})^2$ values are found
for each $l$, we find the average to get $\langle T(\bm{k})^2\rangle$ for each
$l$.  These values are the $C_l$ values of the pure cosmic
string map for all $l$ between $l_{min}$ and $l_{max}$.

Since
\be
T_{G+S}(\bm{k}) \, = \, aT_{G}(\bm{k}) + T_S(\bm{k})
\ee
we obtain
\be
\langle T_{G+S}(\bm{k})^2\rangle \, =  \, a^2\langle T_{G}(\bm{k})^2\rangle+ 
\langle T_{S}(\bm{k})^2\rangle + 2a\langle T_G(\bm{k})T_S(\bm{k})\rangle \, .
\ee
However, $\langle T_G(\bm{k})T_S(\bm{k})\rangle = 0$ because the 
Gaussian and string temperature fluctuations are independent.  Hence
\be \label{equal}
C_{l(G+S)} \, = \, a^2C_{l(G)} + C_{l(S)}.
\ee

Now that we have the $C_l$ values for the simulated CAMB data, 
$C_{l(G+S)}$, the $C_l$'s
we found for the pure cosmic string map, $C_{l(S)}$, and the $C_l$'s for
the simulated gaussian map from CAMB, $C_{l(G)}$, we can compute $a^2$ for
each $l$ by viewing (\ref{equal}) as a defining relation for $a^2$. 
Obviously, for each value of $l$ we will get a different answer. 
To compute a single coefficient $a$ which best fits all of the
$l$'s from $l_{min}$ to $l_{max}$ we perform a linear fit on the coefficients,
$a$, for each $l$.  We use the
best fit $y-$intercept of a linear model fit by minimizing the $\chi^2$
error statistic for $a$.  The y-intercept can be used because the slope of
the best fit to the linear model is negligible.

Once $a$ is found for a particular string image,
the result is averaged over all of the generated simulated string images
to obtain the final value for $a$ which is then used for all the string maps
of a given $G\mu$.

\begin{figure}
\includegraphics[scale=0.8]{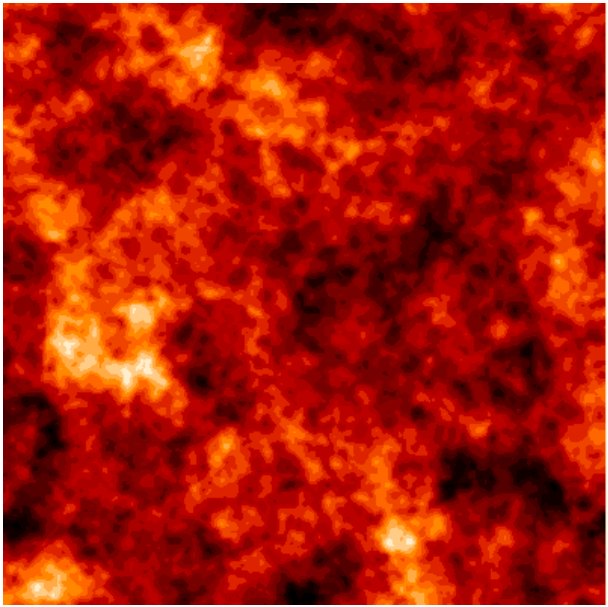}
\caption{The smoothed CMB temperature anisotropy map produced by a
simulation with both Gaussian noise and cosmic strings with
$G \mu =3.5\times 10^{-7}  $ and$N = 10$. In this case
the KS discontinuity lines are visible.} 
\label{combinedmaphigh}
\end{figure}

\begin{figure}
\includegraphics[scale=0.8]{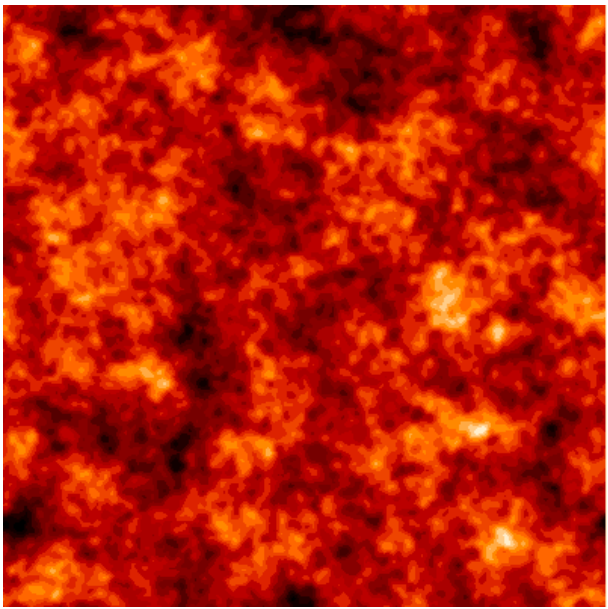}
\caption{The CMB temperature anisotropy map produced by a
simulation with both Gaussian noise and cosmic strings with
a smaller value of $G \mu$, namely $G \mu = 2\times 10^{-8}$ and  $N = 10$. 
In this case the KS lines are not visible by eye.} 
\label{combinedmaplow}
\end{figure}

Figure (\ref{combinedmaphigh}) shows the resulting CMB anisotropy
map in a simulation with a large value of $G \mu$ chosen such
that the strings play an important role. Comparing the map to the
pure string map of Figure (\ref{stringmapN=10}) we see that the
string-induced line discontinuities are still visible. However, in the
case of a lower value of $G \mu$ the effects of the strings are not
visible by eye and we need to resort to a statistical analysis to
study whether the map is distinguishable from that of pure
Gaussian noise (Figure (\ref{combinedmaplow})).

We use the algorithms described in this section to generate maps  with and
without cosmic string signals. The main question we would like to
address now is down to what value of $G \mu$ the Canny edge detection
algorithm is capable of distinguishing between these two sets of maps in
a statistically significant way. In the following section we describe our
realization of the Canny algorithm routine which turns a microwave temperature
map into an edge map, the edges standing for lines across which the gradient
is a local maximum.\footnote{In our realization of the algorithm, we impose an
additional requirement, namely that the gradient magnitude is tailored
appropriately to the expected cosmic string signal.}

\section{Implementation of the Canny Algorithm}

In this section we describe our implementation of
the Canny algorithm.
In its original version \cite{Canny}, the Canny
algorithm is intended to find lines in the map with maximal gradients
across the line. In this way, the algorithm can be applied to the image of
a face and returns a map where only the pixels of the map with the strongest
features are shaded in. Similarly, the algorithm could be applied to find
crystal defect lines on metallic surfaces. Our original work applying the
Canny algorithm to CMB maps \cite{ABB} also was based on this idea.

However, to look for cosmic strings we are less interested in the local maxima of the
gradient map which overall have maximal amplitude. Rather, we are 
interested in local maxima of the gradient for which the amplitude of the
gradient is in correspondence to the expected KS signal. Thus, in contrast
to the original Canny routine which uses two thresholds, we here introduce a modified
algorithm in which three gradient thresholds are made use of (as in \cite{Stewart}).
The use of the thresholds is described below.

We will first give a brief overview of how the algorithm works. 
First, an optional part of the Canny algorithm is to smooth the
map in order to eliminate shot noise. The second step in the algorithm is to
construct a map of the temperature gradients. Next, locations must be
identified which correspond to local maxima of the gradients. The
maximal average gradient of the pure string map determines the
three thresholds used. The fourth part of the algorithm involves
selecting among the grid points identified as local maxima those which
have the right range of magnitudes. Gradients larger than some upper
threshold are not due to single strings and will hence be discarded.
Those above a second threshold quite close to the expected edge
strength will be kept. To take into account the fact that Gaussian noise
may well decrease the amplitude of some pixel along a string edge
below the above-mentioned threshold, we introduce a third (the lowest)
threshold and keep the pixels whose gradient is larger than that
threshold, provided that the pixel is connected to a pixel with gradient
amplitude above the second threshold (and provided that the gradient
directions are appropriate and that the pixel is connected to a pixel in
one of the allowed directions relative to the gradient). 
The pixels thus selected form the Canny
edge map. Given the pixel edge map, a next step in the algorithm is to identify
the edges in the pixel edge map. The length of each edge is found, and a
histogram of edge lengths is produced. The analysis part of the algorithm
then checks if for a fixed value of $G \mu$ the histograms of a pure
Gaussian noise map and of a noise map including a contribution of
cosmic strings (and appropriately reduced amplitude of the Gaussian
noise) are statistically significantly different.

In implementing the Canny algorithm, several (three) choices must be made
beyond fixing the three thresholds used.
We look for the set of choices which give the best differentiation
between maps with and without strings.

The first step in our implementation of the Canny algorithm is the
filtering of the data. This filtering is intended to eliminate point
source noise.  In studies like the present one in which we are dealing
with simulated data which has no noise, the maps can be used
without smoothing. In order to eliminate point sources, smoothing
should be used in the case of real data. We have explored the
effects which smoothing has.

The routine first produces an un-normalized filter map ${\tilde F}(i, j)$ ($i$ and $j$
are integer labels running from $1$ to $N_f$ indexing the
pixels of the filter map)
\be \label{filter}
{\tilde F}(i, j) \, = \, e^{-\frac{x^2 + y^2}{2\sigma^2}}
\ee
in which $x = i - (N_f - 1)/2$, $y = j - (N_f - 1)/2$, $\sigma = 0.5N_f$, 
and $N_f$ is the number of pixels along one direction of the filter map,
an input parameter in the smoothing routine.
The normalization is the sum of all of the unnormalized weights 
\be
C_1 \, = \, \frac{1}{\sum_{ij}{\tilde F}(i,j)} \, .
\ee
Now we can determine the normalized filter map as
\be
F(i,j) \, = \, C_1 {\tilde F}(i,j).
\ee
Finally, we convolve the input temperature map with the filter map to
obtain the filtered data map.  If $M(i, j)$ denotes the initial map, then
the filtered map $FM(i, j)$ is given by
\be
FM(i, j) \, = \, \sum_{k, l = 1}^{N_f} M(i - x, j - y) F(k, l) \, ,
\ee
where $x$ and $y$ are determined from $k$ and $l$ as indicated below
(\ref{filter}).  Boundary points require special treatment. We
repeat the points at the boundary.

%
%

The second part of the algorithm involves constructing the gradient maps.
Two arrays are created, the first containing the magnitude of the gradient
at each pixel of the map, the second containing the information about the
direction of the gradient. 

We compute the lattice derivatives by shifting the temperature map in each 
of the eight directions and looking for the maximal difference.  
Instead of using the usual lattice gradient
which is defined by
\bea
\hat{G}(x,y) \, &=& \, \frac{G(x +\epsilon,y) - G(x-\epsilon,y)}{2\epsilon}\hat{x}   \nonumber \\
&+& \, \frac{G(x,y +\epsilon)-G(x,y-\epsilon)}{2\epsilon}\hat{y} \, ,
\eea
(where the grid spacing is $\epsilon$), 
we shall use the following as our definition of the magnitude of the gradient:
\be
G(x,y) \, = \, \textnormal{max}|\Delta G|
\ee
and the direction is taken to be that for which the maximum 
value of $|\Delta G|$ occurs.  $\Delta G$ is
given by the discrete gradient
\be
\Delta G(x,y) \, = \, \frac{G(n(x,y)) - G(x,y)}{d(x,y)} \, ,
\ee
where $n(x,y)$ denotes the neighboring point to $(x,y)$ and
$d(x,y)$ is the distance between $n(x,y)$ and $(x,y)$.
For neighboring points not on the diagonal, the distance between the 
neighboring point and the point $(x,y)$ is 1, for points along the diagonal 
the distance is $\sqrt{2}$.

The gradient of the image is then stored as an array with a dimension 
containing the
two dimensional pixel array with the edge strengths and a dimension 
containing the
two dimensional pixel array with the edge directions as the values.
  
The next step is to find the pixels of the map which correspond to 
local maxima in the direction of the gradient.
We first shift the map one unit forwards and backwards in each of the 
four directions (along the coordinate axes and along the 
two diagonals, respectively). For each of these four shiftings
we focus on points for which the absolute value of
the gradient is greater than at the two neighboring points
on either side. For all
points selected, we check if the gradient is in the direction of
the shifting. Pixel points which pass this test are kept.  The others
are assigned a value of $0$. Boundary pixels are stripped
from the list of local maxima since their magnitudes are
influenced by boundary effects. In Appendix A we
discuss the local maxima doubling problem which the
code has to deal with.

As mentioned at the beginning of this section, our
implementation of the Canny algorithm makes use of
three thresholds $t_u$, $t_l$ and $t_c$. Their values
are chosen to correspond to the amplitude of the KS effect
of a single string since it is this effect which we want
to identify.

Thus, to set the thresholds for a given value of $G \mu$,
we first run the code on a number of pure string simulations
for that value of $G \mu$. We determine the maximal gradient
(amplitude) $G_{max, i}$ for each simulation, and define the
``string maximal gradient" $G_m$ as the average of the
individual maxima $G_{max, i}$.

Returning to our list of candidate local maxima, then if the
amplitude is larger than $t_c G_m$ the point is
discarded since it is not due to a typical string gradient.
For amplitudes in the range
\be
t_u G_m \,  < \, {\cal A} \, \leq \, t_c G_m
\ee 
then the pixel coordinates and gradient directions are stored in 
arrays. Later on, the edge pixel map will be constructed
by marking all pixel points in the above-mentioned coordinate
array  as $1$. These are pixels for which the gradient is in
the right range to be due to the KS effect from a cosmic string.
Pixels with a slightly lower gradient amplitude might still be
due to the KS effect from a string with amplitude slightly 
reduced by noise. It is advantageous for the success of the
Canny algorithm to take into account this possibility. Thus,
if the amplitude of the local maxima is in the range
\be
t_l G_m \, < \, {\cal A} \, \leq \, t_u G_m
\ee
then the coordinates and gradient directions of such pixels are 
stored in arrays  corresponding to pixel points marked  $1/2$.
Arrays are created to store the indices of the points marked as $1$ and
$1/2$ and separate arrays are created to store the directions of the 
corresponding indices.  

The next step is the ``edgefinder" routine which decides whether
grid points marked as $1/2$ belong to an edge or not. Roughly
speaking, points marked as $1/2$ are considered as belonging
to an edge if they are direct or indirect neighbors to points labelled
by $1$ and the gradients are in a similar direction. To specify
the routine, there are three choices which can be made for this algorithm,
as will emerge below. We start
from a point labelled $1$ (an option in the program is to start
with a point labelled $1/2$ - the first of the
three choices mentioned above)\footnote{ Note that the algorithms starting 
from 1 or 1/2 are
in fact different.  This is also noted in footnote \ref{algorithmdifferences}.}. 
The program then searches in all eight 
directions for a contiguous point labeled as $1/2$ or $1$.  For each 
direction, we check if it lies in the six directions (or 2 directions) perpendicular
or near perpendicular (or just perpendicular - the
second of the choices mentioned above) to the original point's
gradient and check if the gradient is
in the appropriate parallel or near parallel (or just parallel - the third choice)
direction to that at the original $1$ (or $1/2$).  If a point labeled as 1/2 is
found, and the direction is in one of the allowed six (or two), then we
mark the coordinates of that point and repeat the search starting from
that point. We stop if a $1$ is found, in which case we convert all
of the $1/2$s found on the way to $1$. If we do not find another point
labelled $1/2$, we stop and do not change the labelling of the
points encountered on the way.
Then, we move on to the next point labelled by $1$ which has
not already been covered by the search. In the variant of the code
in which the search starts with points labelled by $1/2$, we
search for points labelled by $1/2$ or $1$ as described above, mark
the coordinates of the points found, and continue until either a point
marked $1$ is found (in which case all of the $1/2$ found are
converted to $1$), or else no new point is found, in which case the
$1/2$ are considered not to belong to a common edge and are
not relabelled.
    
The output of the ``edge-finder" routine is a list of coordinates
of the pixels marked by $1$ and of the corresponding gradient
directions.  The result of the algorithm can be represented by
two maps. The first is
a map of pixel points originally labelled as $1$ and $1/2$ (different
shades), the second is an initial ``edge pixel map",
the map of the set of pixels labelled at the end of the
``edgefinder" routine by a $1$.

The final part of the Canny algorithm is an ``edge-counting" routine (this
part of our method goes beyond the standard Canny algorithm).
This routine creates a second ``edge pixel map'' and a 
histogram of number of edges as a function
of the edge length. Starting with the first entry in the list of pixels
marked by $1$, the routine searches in directions perpendicular or
near perpendicular (or exactly perpendicular only) to the gradient direction
to find other pixels labelled as $1$. If such a pixel is found, then
the routine checks if the gradient direction of the new pixel is
near parallel or parallel (or exactly parallel only) to the direction of
the gradient at the initial pixel. If this is the case, the routine
considers the new pixel as part of the edge it is following. 
Once no further pixel is found, the routine considers the edge to
have ended and saves the result.  The program
then moves to the next pixel labelled by $1$. 
Pixels are only allowed to be counted in
one edge.  After all edges are found, the lengths are tallied and can be made
into a histogram.  To take into account
the fact that a pixel may be missing from a string edge due to a
large influence of the Gaussian component, there is an option
for the ``edge-counting" routine to allow for a gap in an edge of
a certain length. This skipping length parameter is a further
input parameter which the user of the algorithm has to set.
As should also be obvious from the discussion in this paragraph,
the same choices of ``perpendicular exactly or nearly perpendicular'' or 
``perpendicular exactly'' and
``parallel exactly or nearly parallel" or ``exactly parallel"
as in the ``edge-finding" routine are open to the user in this part
of the program.  

Let us conclude this section by reminding the reader of the various parameters
which have to be chosen and choices which have to be made in our Canny
algorithm implementation. First of all, in our ``edgefinder"
routine there are the three thresholds $t_c$, $t_u$
and $t_l$ which are used in the labelling of the local maxima. Next, in our
routine to turn $1/2$ pixels into $1$ pixels, there are three choices to be
made, first whether one starts with $1$ or $1/2$ pixels, second whether one
searches for neighboring $1$s and $1/2$s in direction 
perpendicular or in directions
nearly perpendicular or perpendicular to the direction of the gradient, and third whether
one demands that at the selected neighboring site the gradient is parallel
or if it is nearly parallel or parallel to the direction of the gradient at the
initial point. In the edge counting algorithm there are the two choices
analogous to the two choices mentioned at the end of the previous discussion.
The user of the program has similar choices of ``perpendicular exactly
or nearly perpendicular'' versus
``perpendicular exactly'' to the gradient 
for the directions to
search  and ``parallel exactly or nearly parallel"
versus ``parallel exactly'' to make.
Finally, there is the number of points which the ``edge-counting" routine can
skip. 

\begin{figure}
\includegraphics[scale=0.8]{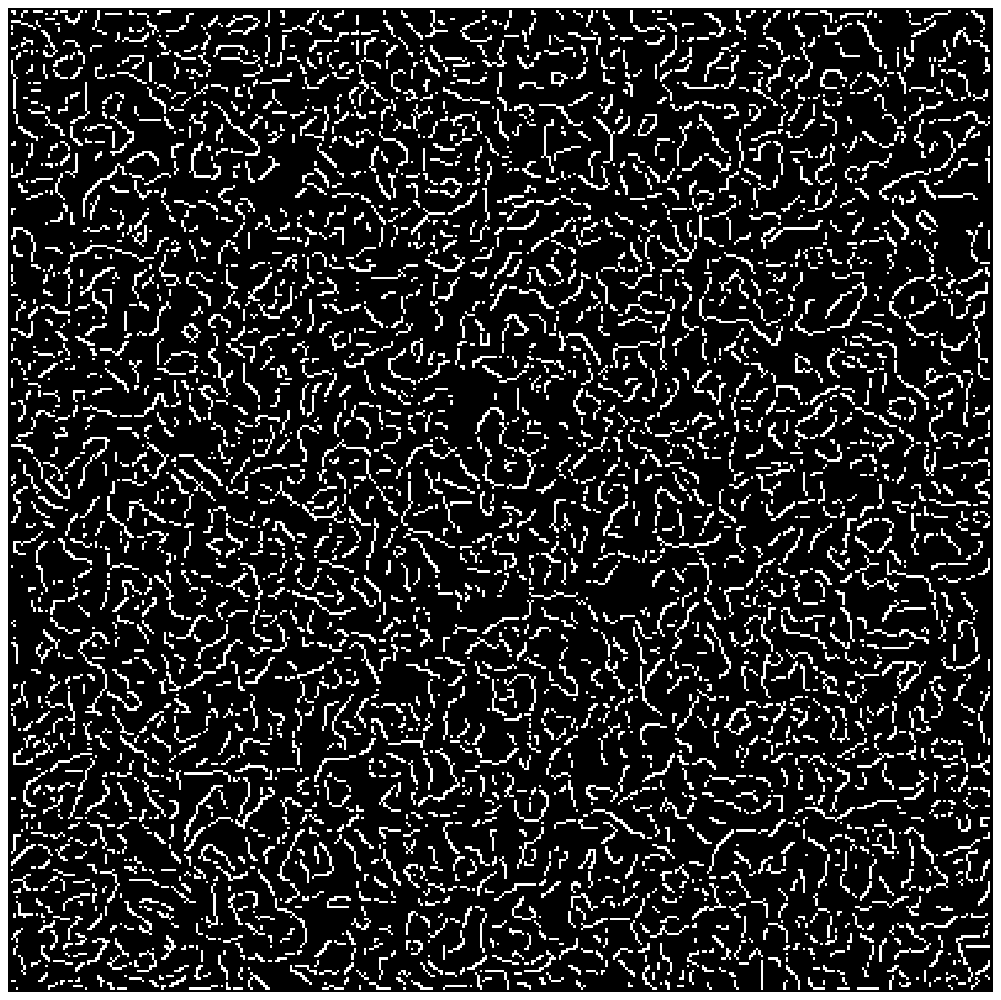}
\includegraphics[scale=0.8]{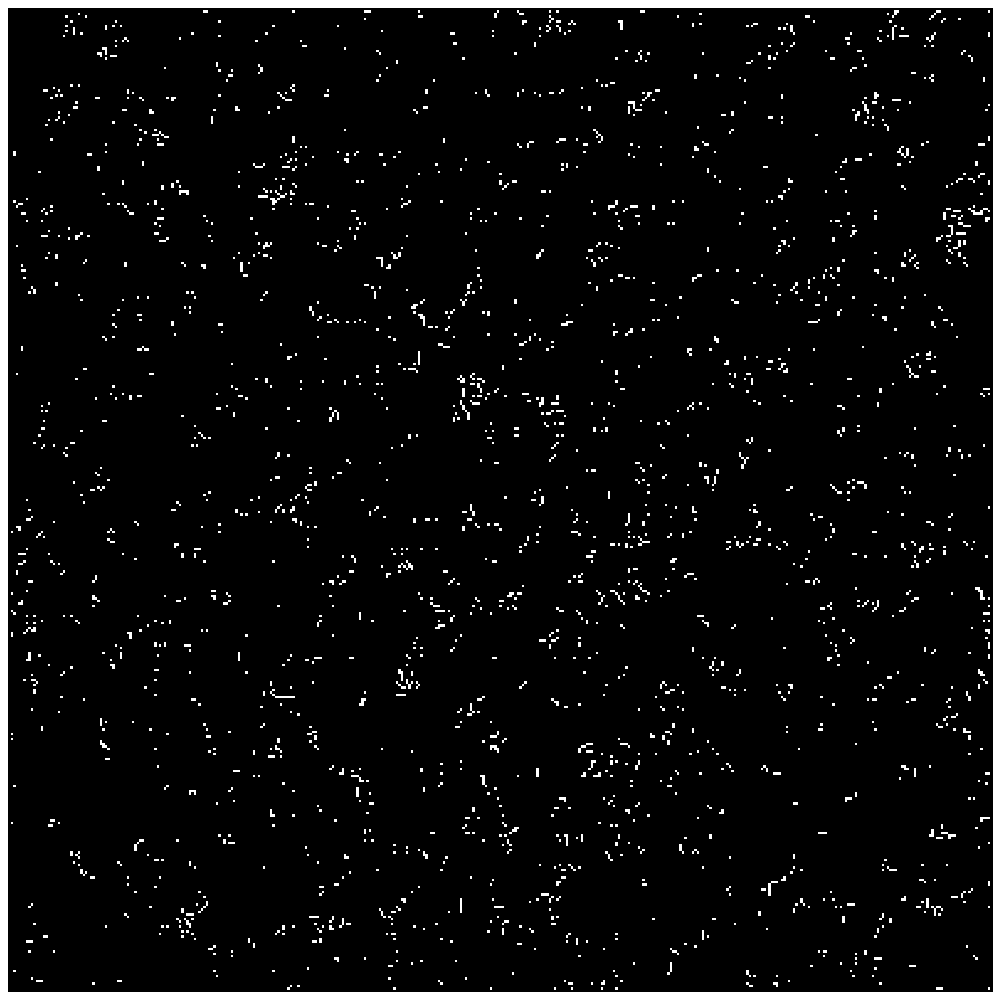}
\caption{Edge maps for the gaussian map shown before.  The first edge map
is for a $G_m$ and thresholds used in the analysis of an $N=10$
gaussian map with strings for $G\mu=3.5\times 10^{-7}$.  The thresholds are
$t_c=2$, $t_u=0.25$, $t_l=0.1$.  The second edge map is made for the
same gaussian map
used in the analysis of an $N=10$ gaussian map with strings for $G\mu=2\times
10^{-8}$.  The thresholds are $t_c=4$, $t_u=0.25$, $t_l=0.1$.}
\end{figure}

\begin{figure}
\includegraphics[scale=0.8]{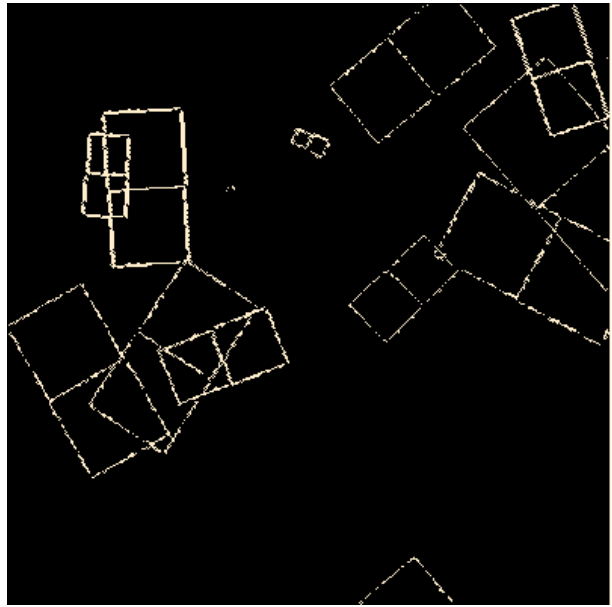}
\caption{The test string edge map.}
\end{figure}

\begin{figure}
\includegraphics[scale=0.8]{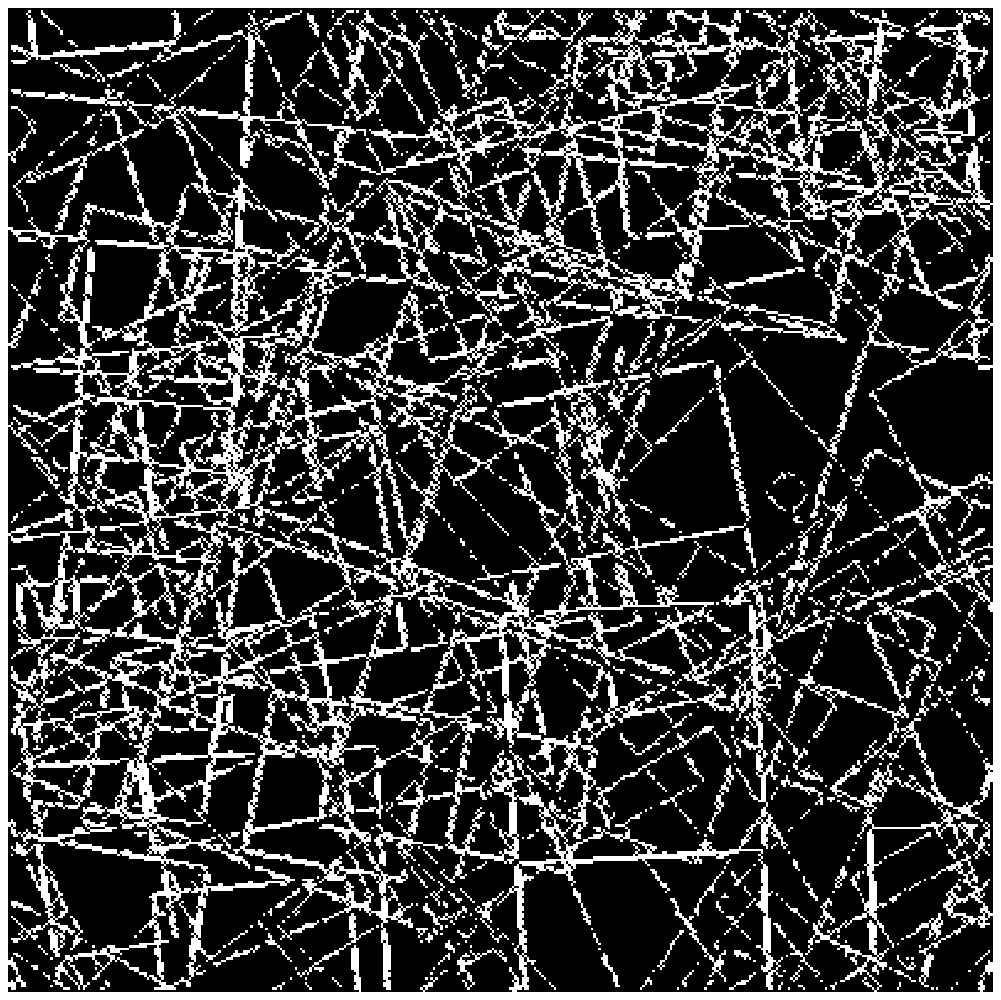}
\caption{N=1 pure string edge map.}
\end{figure}

\begin{figure}
\includegraphics[scale=0.8]{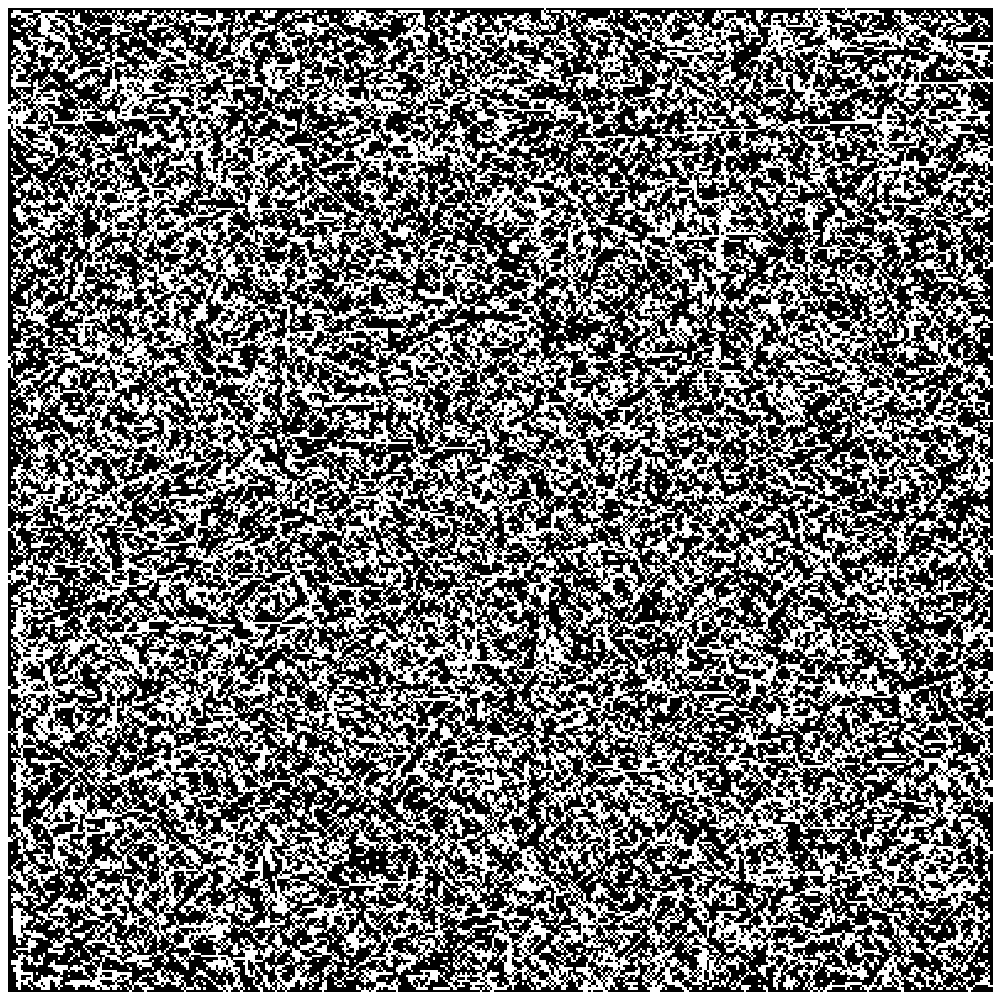}
\caption{N=10 pure string edge map.}
\end{figure}

\begin{figure}
\includegraphics[scale=0.8]{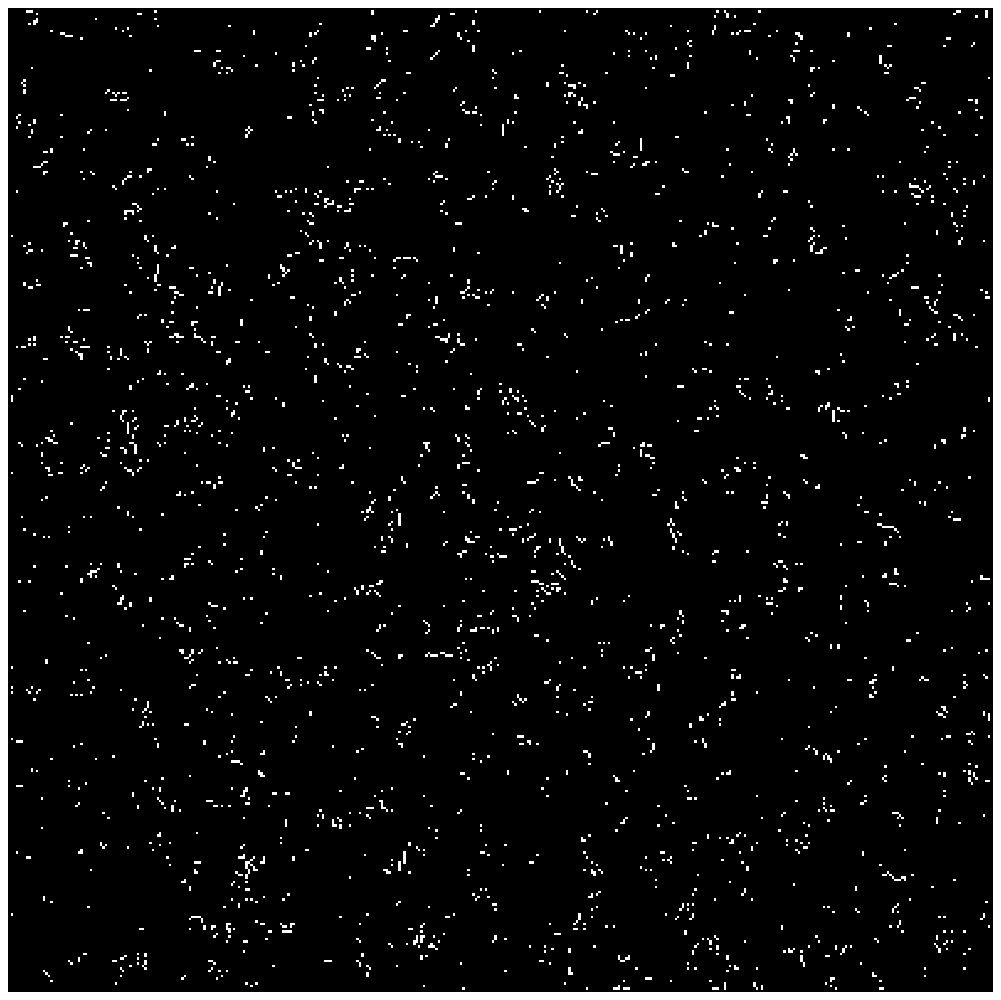}
\caption{$G\mu=2\times 10^{-8}$ unsmoothed string plus gaussian noise edge map. N=10.  The thresholds are $t_c=4$, $t_u=0.25$, $t_l=0.1$.}
\end{figure}

\begin{figure}
\includegraphics[scale=0.8]{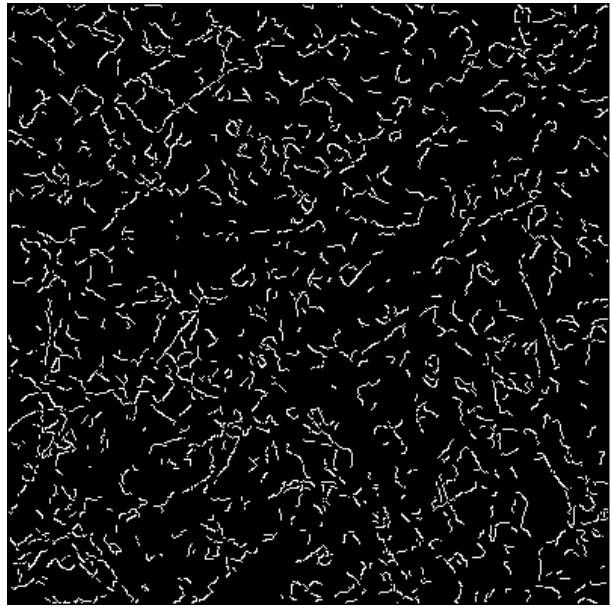}
\caption{$G\mu=3.5\times 10^{-7}$ smoothed string plus gaussian noise edge map.
 N=10.  The thresholds are $t_c=2$, $t_u=0.25$, $t_l=0.1$.}
\end{figure}

\section{Analysis}

We tested the capability of our implementation of the Canny algorithm to
distinguish between temperature anisotropy maps which arise from
a pure Gaussian $\Lambda$CDM model and a model in which there
is a contribution to the power spectrum coming from a scaling distribution
of cosmic strings (and with the power of the Gaussian noise reduced
to maintain the optimal fit to the CMB angular anisotropy power spectrum
as described in Section \ref{simuls}). 

Since we have applications to small-scale CMB anisotropy experiments
in mind, we choose a simulation box of edge length $10^{o}$ and with
angular resolution of $1.5^{'}$. Both in the $\Lambda$CDM model and in
the cosmic string model, the actual universe is a realization of a random
process. Thus, in order to determine whether the Canny algorithm can
distinguish between the predictions of the two theories, we have to run
many realizations of the models. We ran $50$ simulations for both theories.
This number was chosen since it is the number for which the resulting
probabilities converge.

For small values of $G \mu$, we generated each cosmic string map
independently. For larger values of $G \mu$ we generated the temperature
maps by rescaling the temperature map for $G\mu=6\times 10^{-8}$. 
The value of $G_m$ was scaled accordingly.  The value of the scaling
factor $a$ was computed for each of the simulations for fixed value of
$G \mu$ independently. Because of large fluctuations in the various
random variables (e.g. the number of strings), there results a spread
for the values of $a$, and some of the values of $a^2$ were
negative for large values of $G \mu$.  Thus, for large values
of $G \mu$ we used a common value for $a^2$, namely the mean of the 
values of $a^2$ for $G\mu = 6\times 10^{-8}$ and scaled according
to (see \cite{Stewart}):
\bea
a^2 \, &=& \, 1 - \frac{\langle C_l^S\rangle_0}{\langle C_l^G\rangle_0}\left(
\frac{G\mu}{G\mu_0}\right)^2 = 1 + (a_0^2-1)\left(         
\frac{G\mu}{G\mu_0}\right)^2\\
a_0^2 &=& 1- \frac{\langle C_l^S\rangle_0}{\langle C_l^G\rangle_0} \, ,
\eea
where the subscript 0 indicates the reference value of $G \mu$.

Both the models and the Canny algorithm contain a number of parameters.
As described in Section \ref{simuls}, we chose the best fit
parameters of a $\Lambda$CDM model. For the cosmic string distribution,
we considered $G \mu$ to be the free parameter of interest. We fixed
the number of strings per Hubble volume to be $N = 1$ in some
simulations and $N=10$ in other simulations, the string
segment length to be $\gamma H^{-1}$ with $\gamma = 1$, and the
velocity parameter $\tilde{v} = 0.15$.
 
In this work we did a rough optimization of the various parameters which
have to be chosen in the Canny algorithm. For a fixed value of $G \mu$ 
for which the effect of the strings is statistically significant we varied
the parameters to find the parameter values which gave the best
discriminatory power (see below). On this basis, we chose the
various thresholds. Concerning the optimization of the ``flags"
in the routine,  it proved
advantageous to start with the pixels marked $1$ in the edge finding
algorithm, and to look in directions both perpendicular and near to
perpendicular, and allowing the gradient at the neighboring point
to be parallel or near to parallel in both the edge finding and the
edge counting routines \footnote{\label{algorithmdifferences}
We ran the Canny algorithm on two 
different test maps to develop an intuition on how to optimize the flags.  
For both test maps, more edges were found when the number of allowed
directions and gradients were greater.
In one test map more edges were found when the algorithm started with 
points marked as 1/2 and in the other algorithm more edges were found when
the algorithm started with points marked as 1 for some values of the
flags.  Since the algorithm is 
much slower when it starts with points marked as 1/2, in this paper we always
start with points marked as 1.  Longer edges were
found when skipping points was allowed.  This is consistent with the findings
we have for the averages of the full 50 maps of gaussian temperatures and
the combined maps with strings.  Since skipping reduces the number of short
edges, where the standard deviation is smaller compared to the mean, and thus
increases the p-value, it increases the probability that the maps look the 
same.}. 

We ran simulations both with and without smoothing of the maps.
In the runs with smoothing, the smoothing was done in the final
maps (after adding the string maps to the Gaussian maps).
Finally, we ran simulations allowing for skipping of points in
the edge counting algorithm.

Discriminating power was quantified using the {\it Fisher combined
probability test}. This test is applied to the two output histograms
of edge lengths, one from the pure Gaussian simulations, the other
from the strings plus Gaussian maps. For each length $l$, we are
given the mean number of edges of that length and the 
corresponding standard deviation. Given the two means and 
corresponding standard deviations we can apply the t-test
to compute the probability $p_l$ that  the two means come from the 
same distribution. The Fisher combined probability method then 
computes a $\chi^2$ as follows
\be
\chi^2_{2k} \, = \, - 2 \sum_{l = 1}^k ln(p_l) \,
\ee
where $k$ is the number of edge lengths being considered, and computes the
corresponding probability value from a $\chi^2$ distribution with $2k$
values. We chose $k$ to be the last edgelength to have a nonzero 
standard deviation in either the string or gaussian distributions, whichever
is smaller.

The Canny routine makes use of three thresholds.
To find the cutoff or top threshold we ran a script to find the 
maximum average gradient of the fifty string maps and then compared the highest
gradient to the highest average gradient.  This threshold needs to increase
as the $G \mu$ decreases because otherwise one throws out virtually all of the
edges, since as $G \mu$ decreases $G_m$ decreases.  One needs
to keep some edges with
values greater than the maximum gradient in any string map because the 
signal from the gaussian map may add in the same location to the signal 
from the string map. To find the upper and lower thresholds we ran the 
Canny algorithm for various thresholds on a pure string map and determined 
by eye if  enough strings could be seen but not so many that the
overlap of the edges breaks apart the long string edges.
In particular, for $N = 10$ if the thresholds are too small the images are
saturated with short edges which break up the long edges.  Thus,
the thresholds need to be higher for $N = 10$ than for $N = 1$.
We leave a detailed optimization of
the thresholds to future work.

As a test of the code, we ran the analysis algorithm on two independent sets
of fifty Gaussian CMB maps and computed the probability that the two sets
come from the same Gaussian ensemble. The resulting probability was
typically of the order of $0.7$, i.e. within the $1 \sigma$ error range.

\begin{figure}
\includegraphics[scale=0.5]{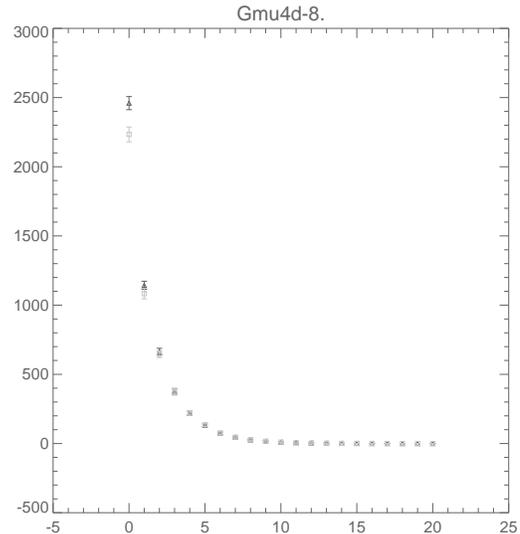}
\caption{Histogram of the edge length distribution of the unsmoothed 
pure Gaussian
maps and the strings plus smoothed Gaussian maps for a value of 
$G \mu = 4\times 
10^{-8}$.  The number of edgelengths for strings plus gaussian maps are denoted
by triangles and the squares denote the number of edgelengths for the pure
gaussian maps.  We plot the means and the standard deviation as the error.
Note the systematically larger number of edges for maps with strings for short
edgelengths.}
\label{thist1}
\end{figure}

Our simulations show that maps originating from cosmic strings have
a larger number of edges than the corresponding pure
Gaussian maps. This is a consequence of the presence of the Kaiser-Stebbins
edges, the signal we are looking for. Figure \ref{thist1} shows a comparison
of the two histograms for the value $G \mu = 4\times 10^{-8}$. Since the Gaussian
noise cuts up the long edges, the difference in edge lengths is statistically
significant only for short edges. 

Table 1 summarizes our results obtained by applying the Fisher combined
probability method to the histograms of string edge lengths. The results
are for $N = 10$ ($10$ strings per Hubble volume per Hubble time), for
un-smoothed maps, without skipping in the edge drawing and edge
counting routines. The statistical analysis includes edgelength 1 data.
The first column gives the value of $G \mu$, the second the value of
the top threshold (the other two thresholds are held fixed at the values
$t_u = 0.25$ and $t_l = 0.1$), the third gives the probability that
the histograms of the string and Gaussian maps come from the same
Gaussian distribution. The fourth value gives the maximal edge length
used in the analysis, and the last column lists the value of $a$. We see that the cosmic 
string signal can be detected \footnote{More conservatively, we should
say that the difference between the maps with cosmic strings and pure Gaussian
maps can be detected.} at a three sigma level
down to a value $G \mu = 2 \times 10^{-8}$ which is almost two
orders of magnitude better than the current bounds obtained by direct
searches, and one order of magnitude better than the limits
on $G \mu$ coming from the CMB angular power spectrum  constraints.
These results are for unsmoothed maps. 

As can be expected from looking at the histograms of \ref{thist1}, the
sensitivity of the algorithm decreases slightly if edges of length 1 are excluded
from the analysis. The results excluding edges of length 1 are shown
in Table 2.

Since smoothing dramatically reduces the number of short edges, and since
the strength of the signal of our analysis comes from the number of short edges,
it turns out that smoothing significantly weakens the ability of the algorithm to 
pick out strings. The results of our analysis applied to smoothed maps (smoothing
length 3) are given in Table 3. The loss in discriminatory power is about a factor
of 5. This is a serious concern when considering applications of our algorithm
to real data. 

In Table 4 we present the results for $N = 1$. Since there are an order of magnitude
less string edges in this case, the limit on $G \mu$ which can be obtained is slightly
weaker  (about a factor of 3 weaker). To put this result into context, it is important to
point out that limits on $G \mu$ from other studies implicitly or explicitly use $N = 10$.
The limits on $G \mu$ from matching the angular CMB power spectrum would be
weaker by about one order of magnitude.

The results of the previous tables were obtained from an algorithm which 
did not have any skipping in the edge counting routine. We allowed skipping
of two points. Skipping of two points leads to a larger number of long edges.
Skipping by four points leads to edge broadening, an unwanted feature.
Hence, we only considered skipping two points. Including
skipping turns out to reduce the number of short edges more significantly
than it increases the number of longer edges. Thus, the power
of our routine to discriminate between maps with and without strings
slightly decreases when introducing skipping (which is the opposite
of what we initially expected). The results are indicated in the following
table. However, it is possible that further optimization of the routine
parameters would reverse the results concerning the effectiveness
of skipping.

\begin{widetext}

\begin{table}[h]
Table 1: Unsmoothed String Map versus Gaussian Map N=10\\
Minimum Edge Length: 1\\
$t_u=0.25$, $t_l=0.1$, num. skipped points=0\\
\begin{tabular}{|c|c|c|c|c|}
\hline
Gmu&   $t_c$ & probability &  max. edgelength&mean a\\
\hline
   1.0000000e-08 &     7.00000 &     0.44757170 &      11.000000 &     
0.99967664\\
\hline
   2.0000000e-08 &     4.00000&    0.0020638683&       14.000000 &0.99862559\\
\hline
   4.0000000e-08 &     3.00000&       0.0000000  &     21.000000&  0.99447207\\
\hline
   6.0000000e-08 &     3.00000 &      0 &     33 &     0.98754092\\
\hline
\end{tabular}
\end{table}

\begin{table}[h]
Table 2: Unsmoothed String Map versus Gaussian Map: N=10\\
Minimum Edge Length: 2\\
$t_u=0.25$, $t_l=0.1$, skipped points=0\\
\begin{tabular}{|c|c|c|c|c|c|}
\hline
Gmu&   $t_c$ & probability &  max. edgelength&mean a\\
   1.0000000e-08 &     7.00000 &     0.56022527 &      11.000000  & 0.99967664
\\
\hline
   2.0000000e-08 &     4.00000 &    0.087458606 &      14.000000 & 0.99862559\\
\hline
   4.0000000e-08&      3.00000&   4.8140603e-11&  21.000000&    0.99447207\\
\hline

   6.0000000e-08 &     3.00000 &      0.0000000  &     33.000000 & 0.98754092\\
\hline
\end{tabular}
\end{table}

\begin{table}[h]
Table 3\\
Minimum Edge Length:       1\\
Smoothed String Map versus Gaussian Map N=10\\
$t_u=0.25$, $t_l=0.1$, num. skipped points=0\\
\begin{tabular}{|c|c|c|c|c|c|}
\hline
Gmu &  $t_c$ & probability &  max. edgelength &  mean a\\
\hline
   6.0000000e-08&      3.00000&      0.10701445&       42.000000&  0.98754092\\
\hline

   8.0000000e-08 &     3.00000& 0.059179809 &      46.000000&   0.97932855\\
\hline
   9.0000000e-08 &     3.00000&   0.00012533417&       46.000000&      0.97174247\\
\hline

   1.0000000e-07 &     3.00000&   2.4759972e-11&    52.000000&    0.96499435\\
\hline
   1.5000000e-07&      2.50000 &      0.0000000 &      59.000000 &0.91936485\\
\hline

\end{tabular}
\end{table}

\begin{table}[h]
Table 4\\
Minimum Edge Length:      1\\
Unsmoothed String Map versus Gaussian Map: N=1\\
$t_u=0.03$, $t_l=0.005$, skipped points=0\\
\begin{tabular}{|c|c|c|c|c|c|}
\hline
Gmu &  $t_c$ & probability &  max. edgelength&mean a\\
\hline
   2.0000000e-08&      4.00000&      0.43985131&       7.0000000& 0.99986506\\
\hline
   4.0000000e-08&      3.00000&      0.76437603 &      11.000000&  0.99937157\\
\hline
   6.0000000e-08 &     3.00000&    0.0019923380&  19.000000&      0.99901436\\
\hline
   8.0000000e-08&      3.00000&   8.7787192e-09&  28.000000&      0.99816180\\
\hline
   9.0000000e-08 &     3.00000 &  3.8369308e-13&       32.000000& 0.99757307\\
\hline
\end{tabular}
\end{table}

\begin{table}[h]
Table 5\\
Minimum Edge Length:       1\\
Unsmoothed String Map versus Gaussian Map:N=10\\
$t_u=0.25$, $t_l=0.1$, skipped points=2\\
\begin{tabular}{|c|c|c|c|c|}
\hline
Gmu &  $t_c$&  probability &  max. edgelength&mean a\\
\hline
   1.0000000e-08&      7.00000&      0.43783402&       41.000000&      0.99967664\\
\hline

   2.0000000e-08 &     4.00000 &     0.30944903 &      55.000000& 0.99862559\\
\hline
   4.0000000e-08&      3.00000&   5.4179907e-08&       139.00000 &     0.99447207\\
\hline

\end{tabular}
\end{table}

\end{widetext}

\section{Conclusions and Discussion}
 
 We have developed a new program to search for cosmic strings in CMB anisotropy maps
 making use of the Canny algorithm, and have tested it with simulated data corresponding
 to maps with specifications corresponding to those of current ground-based CMB
 experiments. The code contains a number of optimization parameters, and we have discussed
 the role of these parameters.
 
 Our results (based on a rough optimization of the parameters) show that our algorithm
 has the potential to improve the bounds on the cosmic string tension from direct CMB
 observations by up to two orders of magnitude compared to existing limits, and by
 one order of magnitude from other more indirect limits.  The limiting
value of $G\mu$ decreases as $N$ (the number of strings per Hubble volume per
Hubble time) increases.  For cosmic superstrings the intercommutation
probability may be much lower than for Abelian field theory strings, thus
leading to a larger value of $N$ \cite{Pol}.  Hence, our method may
be able to set more stringent limits of the value of $G\mu$ in the case of 
cosmic superstrings.
\footnote{While this manuscript was being prepared for submission, a
preprint \cite{Lausanne} appeared in which a novel method different
from ours was proposed to search for cosmic strings. With this new
method, improved limits on the cosmic string tension also appear to
be possible.}

A possible concern with our analysis is that our toy model of
representing the effects of a string on the microwave sky
introduces artificial edges. Although the presence of the
central edge in the CMB pattern of a string is clearly physical,
the edges perpendicular to this central edge are due to
the toy model which considers finite length string segments
rather than a network of infinite strings. Although
the temperature perturbation clearly goes to zero at a
perpendicular distance $t$ from a string present at time $t$,
the sharpness of the change given by \cite{Joao} may be
modulated by interactions between the time of formation of
the string network and the time $t$ \footnote{We thank the
Referee for raising this issue.}. Thus, one of our
simulations with $N$ strings per Hubble volume per Hubble
time may contain as many edges as a ``real'' string map
with $4N$ strings per Hubble volume per Hubble time.
As can be seen by comparing Tables 4 and 5, the dependence
of the limits on $G \mu$ on $N$ is not large. The limit
on $G \mu$ appears to scale as $N^{-1}$. Thus, the limit
on $G \mu$ obtained from our analysis might be a factor
of $3$ too optimistic due to the larger number of edges.
This point could also be addressed by a Canny algorithm analysis of
more sophisticated string simulations, which we shall pursue as a 
follow-up study.

The results presented here are based on comparing simulations with and without
cosmic strings of a known value of $G \mu$. However, we do not know the
value of $G \mu$, and the goal of observations is to determine the value or to
set limits on it. Hence, the way we plan to apply our algorithm to real data
is to compare the edge histograms based on real data with those of a cosmic
string plus Gaussian map with fixed $G \mu$, and varying $G \mu$, starting
from the current limit for this quantity. Limits on $G \mu$ can be obtained
if the histogram derived from the data is statistically different from that of
a Gaussian plus strings map for the corresponding value of the normalized
string tension. A candidate detection of cosmic strings would require the
histogram derived from the real data to be statistically indistinguishable 
from that of a Gaussian plus strings map, and at the same time different from 
that of a pure Gaussian map.  If there are indeed cosmic strings in the sky 
with a value
of $G \mu$ for which our analysis gives a probability $p$ that
the histograms with and without strings are from the same
distribution, then the probability that our statistical analysis
would give a positive detection of these strings will be
$1 - p$.

When applying the algorithm to real data, a serious concern is systematic errors 
 introduced by the observing strategy - specifically lines in the maps due to the
 scanning strategy. In specific experiments such as the South Pole Telescope (SPT)
 experiment, lines introduced by the scanning will be in one specific direction,
 and thus the algorithm can be rendered immune to this effect by considering
 only edges which are not parallel to the direction of the scanning stripes. Foreground
 and instrumental noise are other important problems. Smoothing of the maps is 
 supposed to reduce the latter
 problem, and the study of \cite{Stewart} confirms that instrumental noise such
 as that anticipated in the SPT telescope will not have a big effect on the power
 of the algorithm. 
 
 The next step of our research program will be to apply our code to existing data.
 The main challenge will be to take care of the systematic errors contained in
 the data. We plan to compare data for the actual sky map to Gaussian
 and Gaussian plus strings simulations. In order to put the data in a format for
 which a statistical analysis is possible, we will divide the sky maps into sub-maps
 of equal size, and treat each sub-map as a different realization of the sky. In
 other words, we make the ergodic hypothesis and identify ensemble and
 spatial averaging. Thus, if we want to compare squares of edge length
 $10^{0}$ in the sky, then we need a total survey area $50$ times larger
 in order to be able to have $50$ independent data sets for the sky.
 We also plan to apply our code to more realistic cosmic string simulations,
 such as those of \cite{Fraisse}.
 
\begin{acknowledgments} 
 
This work is supported in part by a NSERC Discovery Grant and by funds from
the CRC Program. We are grateful to Stephen Amsel, Andrew Stewart and
Eric Thewalt for many discussions on numerical issues.
We also wish to thank Matt Dobbs, Aurelien Fraisse,
Christophe Ringeval, and in particular Gil Holder for useful communications.
R.D. would like to thank Andrew Frey for useful discussions.
\end{acknowledgments} 
 
\section{Appendix A}
 
In this Appendix we describe the maxima doubling problem which arises
in our implementation of the Canny algorithm. Let us illustrate the
problem in terms of an example.

If the temperature grid points are
\be
\begin{matrix}
6 & 8 & 10 & 20 \\
3 & 5 & 12 & 14 \\
4 & 6 & 9 & 11
\end{matrix}
\ee
and the points are labeled starting from 0 from left to right top to bottom,
then the point at grid point 5 (temperature 5) is a local maximum 
with gradient 7 along the
0 direction.  The grid point 6 (temperature 12) is a local maximum along the
1 direction with gradient 8.  There aren't any doubles in this case.  

However, if we had
\be
\begin{matrix}
6 & 8 & 10 & 15 \\
3 & 5 & 12 & 14 \\
4 & 6 & 9 & 11
\end{matrix}
\ee
then the grid point 5 (temperature 5) is a local maximum along the 0 direction
with a gradient of 7 and the grid point 6 (temperature 12) is a local maximum
along the 4 direction with gradient 7.  We use a sorting routine to remove the
first occurrence of a gradient of 7 (index 5) as we want there to be only one
local maximum at this point.

\end{document}